\documentstyle[twocolumn,floats,aps,epsf,prb]{revtex}
\begin{document}
\draft
\title{The $c$ Axis Optical Conductivity of Layered Systems in the 
Superconducting State}
\author{W. A. Atkinson\cite{bill} and J. P. Carbotte}
\address{Department of Physics and Astronomy, McMaster University,
Hamilton, Ontario, Canada L8S 4M1}
\date{\today}

\maketitle
\begin{abstract}
In this paper, we discuss the $c$ axis optical conductivity 
Re $[\sigma_c(\omega)]$ in the high $T_c$ 
superconductors, in the superconducting state.  
The basic premise of this work is that electrons 
travelling along the $c$ axis between adjacent CuO$_2$ layers
must pass through several intervening layers.  In earlier work
we found that, for weak inter-layer coupling, 
it is preferable for electrons to travel along the $c$ axis by 
making a series of interband transitions rather 
than to stay within a single (and very narrow) band.
Moreover, we found that many of the properties of the normal state
optical conductivity, including the pseudogap could be explained by 
interband transitions.
In this work we examine the effect of superconductivity on the interband
conductivity.  We find that, while the onset of superconductivity
is clearly evident in the spectrum, there is no clear signature of
the symmetry of the superconducting order parameter.
\end{abstract}


\section{Introduction}
\label{secI}
Recently, there has been a lot of attention paid to the $c$ axis optical
conductivity in the high $T_c$ cuprate superconductors.  In particular,
measurements of the optical conductivity in YBa$_2$Cu$_3$O$_x$
(YBCO$_x$) have revealed that the $c$ axis transport is very different
in character from the electronic transport in the $a$ and
$b$ directions (within the layers).\cite{homes,schutzmann}   
There is a great deal of speculation
as to the source of this difference.  At one end of the spectrum of
thought there is the claim that the unusual $c$ axis transport is evidence
for some non-Fermi liquid like ground state within the CuO$_2$ layers.%
\cite{anderson}
At the other end of the spectrum, the claim is that the electronic
ground state is metallic, but that there is some unconventional
tunneling mechanism between the layers.\cite{zha,graf,hirschfeld,%
rojo,leggett,kumar}  The common feature to all
of these models, however, is that they consider only a 
single copper oxide layer in each unit cell.

In the case of YBCO$_x$, of course, the situation is not so simple.
There are several layers between adjacent CuO$_2$ layers, one of
which (the CuO chain layer) is known to be conducting.  In a previous
paper\cite{atkinsonI}
we asked the question ``What is the effect of these intermediate
layers, assuming a simple metallic model?''  We considered a simple
two-layer model in which each unit cell  contained a CuO$_2$ plane
and a CuO chain, and we calculated the optical conductivity in the
normal state.  What we found was that, while making almost no assumptions
about the band structure, we could explain many of the features seen in
experiment---including the pseudogap seen in optical experiments.
We claim that this work throws into doubt any attempt to interpret 
the $c$ axis conductivity that does not take into account the 
multilayered structure of the high $T_c$ materials.  

In this work, we examine the $c$ axis optical conductivity for a simple
model of YBCO$_x$ in the superconducting state.  This model is the
same as the one described above:  each unit cell contains a plane layer
and a chain layer.  These layers are evenly spaced and connected by 
coherent single-electron hopping.  The amplitude for the hopping is
parameterized by $t_\perp$.  The current model differs from our earlier
one in two important ways.  First, the sample is taken to be superconducting.
Based on experimental observations, we include a superconducting gap
in both the plane and chain layers.  \cite{atkinsonII}
Second, the scattering rate  $1/\tau$ is set to 
zero. In our study of the optical conductivity in the normal state, the linear
temperature dependence of the scattering rate was important at high
temperatures.  At
low temperatures the scattering rate is small and can be ignored.

One of the main conclusions of our earlier work is that, for small
$t_\perp$, interband processes play a dominant role in the $c$ axis
conductivity.  In isolation, the plane and chain
layers have dispersions $\xi_1(k_x,k_y)$ and $\xi_2(k_x,k_y)$
respectively.  When the layers are
coupled by $t_\perp$, they hybridize and form two bands 
$\epsilon_+(k_x,k_y,k_z)$
and $\epsilon_-(k_x,k_y,k_z)$.  Electrons which travel along the $c$ axis 
in the presence of an external field may
do so by either staying within the bands $\epsilon_\pm$, or by
making transitions between them.  If the layer dispersions $\xi_1$
and $\xi_2$ are nondegenerate, then the bands $\epsilon_\pm$ will
differ from $\xi_1$ and $\xi_2$ by 
$\sim (2t_\perp \cos(k_zd/2))^2/(\xi_1-\xi_2)$, where $d/2$ is the interlayer
spacing (this follows from
second order perturbation theory).  The Fermi velocities $v_{z\pm} = 
\hbar^{-1} \partial \epsilon_\pm/\partial k_z$ therefore scale as $t_\perp^2$.
On the other hand, the matrix element for an interband transition
is  $\sim 2t_\perp \cos(k_zd/2)$.  For small $t_\perp$, the $c$ axis transport
occurs preferentially through interband transitions.

While the intraband conductivity has the well-known Drude form, the
interband conductivity does not.  Much of this paper will be devoted
to understanding how the interband conductivity in the plane-chain model
differs from the usual picture of conductivity in the superconducting state.

We are aware of one other model which attempts to account for the presence
of intermediate layers.  The model, proposed by Abrikosov,\cite{abrikosov}
examines the effect of resonant tunneling through an impurity layer which lies
between CuO$_2$ layers. In the case of YBCO$_x$, the impurities
are the oxygen vacancies in the CuO chain.  With his model, he is able
to provide a reasonable explanation of the $c$ axis d.c.\ resistivity.

This paper is organized as follows.  In Sec.~\ref{secII}, an
equation for $Re[\sigma_c(\omega)]$ (the optical conductivity along 
the $c$ axis) is derived.  The equation is integrated numerically
and the results are discussed in Sec.~\ref{secIII}.  A brief conclusion
is contained in Sec.~\ref{secIV}.

\section{Derivations}
\label{secII}

The mean field Hamiltonian for our model is\cite{atkinsonII}
\begin{equation}
H = \sum_{\bf k} C^\dagger_{\bf k} \, h({\bf k}) \, C_{\bf k}
\end{equation}
where $C^\dagger_{\bf k} = [c^\dagger_{1{\bf k}\uparrow},
c_{1-{\bf k}\downarrow},
c^\dagger_{2{\bf k}\uparrow},c_{2-{\bf k}\downarrow}]$ and 
$c^\dagger_{i{\bf k}\sigma}$
creates an electron of spin $\sigma$ and 3-dimensional momentum 
${\bf k}$ in the plane
($i=1$) or chain ($i=2$) sublattices.  The Hamiltonian matrix is
\begin{eqnarray}
h({\bf k}) = \left [
        \begin{array}{cccc}  \xi_1 & \Delta_{\bf k} & t({\bf k}) & 0 \\
                            \Delta_{\bf k} & -\xi_1 & 0 &-t({\bf k}) \\
		            t({\bf k}) & 0 &  \xi_2 & \Delta_{\bf k} \\
			    0 &-t({\bf k}) & \Delta_{\bf k} & -\xi_2 
	\end{array}
\right],
\label{9}
\end{eqnarray}
where $\Delta_{\bf k}$ is the mean field superconducting order parameter,
$\xi_1$ and $\xi_2$ are the plane and chain dispersions respectively,
and $t({\bf k})$ couples the plane and chain layers.
The dispersions $\xi_1$ and $\xi_2$ are 
\begin{equation}
\xi_1 = - 2\sigma_1 [\cos(k_xa)+\cos(k_ya) - 2B \cos(k_xa)\cos(k_ya)]-\mu_1,
\label{1}
\end{equation}
and
\begin{equation}
\xi_2 = - 2 \sigma_2 \cos(k_ya) - \mu_2
\label{2}
\end{equation}
where $\sigma_1,\,\sigma_2,\,\mu_1,\,\mu_2$, and $B$ are adjustable
parameters, and $a$ is the unit cell size in $ab$ directions.  
For this work, the parameters are fixed at $\{\sigma_1,\sigma_2,\mu_1,
\mu_2\} = \{70,100,-65,-175\}$ meV and $B=0.45$.  This is done
by fitting the magnitudes of the penetration depth at $T=0$ (which effectively
measures the magnitudes of the Fermi velocities $v_x$, $v_y$ and $v_z$)
to experiment,\cite{basov} 
while maintaining a Fermi surface that looks qualitatively
like that of band structure calculations.\cite{pickett,andersen,jyu}

The interlayer coupling is 
$t({\bf k}) = -2t_\perp \cos(k_z d/2)$, which follows from a 
tight-binding model of $c$-axis coupling.  The unit cell size is $d$ along the
$c$-axis.  Again, by fitting $\lambda_c(T=0)$ to experiment, we have 
determined that $t_\perp \sim 20$ meV for YBCO$_{6.93}$ and 
$t_\perp \sim 1$ meV for YBCO$_{6.7}$.

\begin{figure}[p]
\begin{center}
\leavevmode
\epsfxsize .8\columnwidth
\epsffile{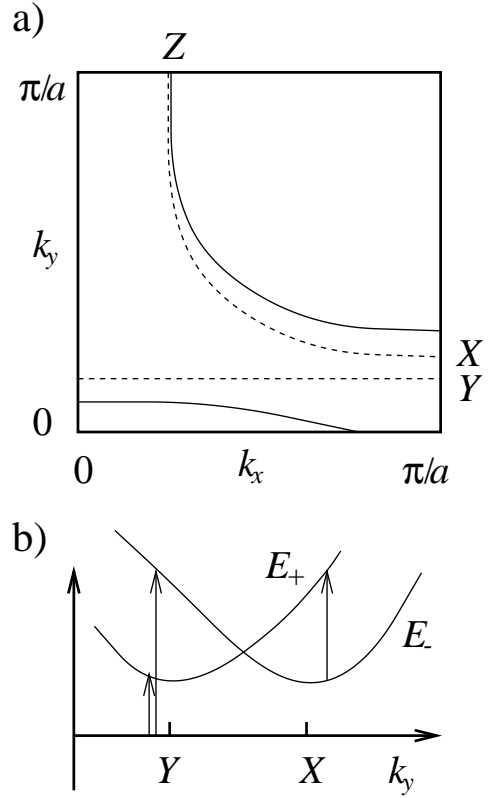}
\caption{a) Fermi surface for the plane-chain model.  The dashed lines
are the Fermi surfaces of the isolated plane and chain layers, the
solid lines are the Fermi surfaces at $k_z = 0$.  For other values
of $k_z$, the Fermi surfaces lie between the solid and dashed curves.
b)  Schematic of the energy bands along $k_x = \pi/a$, $k_z = \pi/d$.
The energy bands are shown for the superconducting state
(solid curves).  The normal state energies $\epsilon_-$ and
$\epsilon_+$ vanish at the  points $X$ and $Y$ respectively.
External fields excite two types of interband
transition---pair creation and quasiparticle transitions---which are shown.
In pair creation, the final state has one quasiparticle in each of the
bands.  This process is gapped since the energy required for the process
$\hbar \omega = E_+({\bf k}) + E_-({\bf k})$ has a nonzero minimum.  
In quasiparticle
transitions, thermally excited quasiparticles can then make interband
transitions.  This process is not gapped since there will generally be
values of ${\bf k}$ for which the excitation energy $\hbar \omega = 
E_+({\bf k}) - E_-({\bf k})$ vanishes.}
\label{f4a}
\end{center}
\end{figure}

The Fermi surface is shown in Fig.~\ref{f4a}(a).  The dashed lines give the
Fermi surface of the isolated chain and plane layers.  These are also
the Fermi surfaces of $\epsilon_\pm$ at $k_z = \pi/d$, since $t(\pi/d)=0$.
The solid curves are the Fermi surfaces at $k_z=0$, where $t(k_z)$ is
a maximum.  For intermediate values of $k_z$, the Fermi surface lies
between these two curves.

The superconducting gap $\Delta_{\bf k}$ is chosen to be the same in
both the plane and chain layers.  It can have either a $d$-wave
symmetry
\[
 \Delta_{\bf k} = \Delta(T)[\cos(k_xa) -\cos(k_ya)],
\]
or an $s$-wave symmetry
\[
 \Delta_{\bf k} = \Delta(T).
\]
In this work we do not propose a microscopic origin for the pairing
interaction but rather assume that $\Delta(T)$ can be described
phenomenologically by
\[
\frac{\Delta(T)}{\Delta(0)} 
= \mbox{tanh} \left( \frac{\Delta (T)}{\Delta(0)} 
\frac{T_c}{T}\right) 
\]
with $2 \Delta(0)/k_BT_c = 3.5$ and $T_c = 100$ K.

We have found, in previous work,\cite{atkinsonI} 
that Re $[\sigma_c(\omega)]$ in the normal
state depends on the choice of parameters in two ways: $\sigma_c(\omega)$
depends qualitatively on whether or not the chain and plane Fermi surfaces
cross, and on the magnitude of $t_\perp$.  We found that when the 
Fermi surfaces do not cross, there is a direct gap which appears as a 
pseudogap of energy $\hbar \omega_0$ in $\sigma_c(\omega)$.  
The pseudogap disappears at high $T$ because
of the smearing of the quasiparticle energy by the large, temperature
dependent, scattering rate.  The temperature $T^\ast$ at which the
pseudogap begins to be resolved is therefore determined by
\[
1/\tau(T^\ast) \sim \omega_0,
\]
and there is no direct connection between the onset of superconductivity
at $T_c$ and $T^\ast$.
As we mentioned before, the scattering is
ignored in this work so the pseudogap is unchanged by temperature.  We also
found in our earlier work that for small values of $t_\perp$ the
Drude-like intraband contribution to the conductivity is hidden by the large
interband conductivity.

In this work we will examine the $c$-axis optical response in the 
superconducting state.  We will assume that the sample is in the clean
limit.  There is good evidence\cite{Hardy} that the large, temperature 
dependent
scattering rate does, in fact, drop dramatically below $T_c$, although
we have made this assumption primarily to keep the model simple.
In this case, the Drude part of $\sigma_c(\omega)$ collapses to
a $\delta$-funtion at $\omega=0$, and the only response at finite
frequency comes from interband transitions.  Since the main effect
of changing $t_\perp$ is to change the magnitude of the interband
conductivity and not its form,  we can arbitrarily fix $t_\perp = 20$ meV.

One of the most interesting features of our model Hamiltonian is that
the gap $\Delta_{\bf k}$ is the same in both the chain and plane layers.
We have discussed the experimental evidence for this elsewhere.%
\cite{atkinsonII}
It is a somewhat surprising property since the chain layer does not
have the tetragonal symmetry needed to generate a $d$-wave order parameter.
This strongly suggests that the pairing interaction must originate in
the (tetragonal) CuO$_2$ plane layer, but then the difficulty lies in
understanding the large magnitude of the gap on the chain layer.
This has been discussed elsewhere by us,\cite{atkinsonII,atkinsonIII}
by Xiang and Wheatley\cite{xiang} and
by O'Donovan and Carbotte.\cite{odonovan}  

The eigenvalues of the Hamiltonian matrix $h({\bf k})$ give the band 
energies
$E_1=E_+$, $E_2=-E_+$, $E_3 = E_-$, $E_4=-E_-$, where
\begin{equation}
E_\pm = \sqrt{\epsilon_\pm^2 + \Delta_{\bf k}^2}
\end{equation}
and $\epsilon_\pm$ are the normal state band energies
\begin{equation}
\epsilon_\pm = \frac{\xi_1+\xi_2}{2} \pm \sqrt{\left[\frac{\xi_1-\xi_2}
{2}\right]^2 +t^2}.
\label{13}
\end{equation}

The optical conductivity for a layered system is\cite{atkinsonII}
\begin{eqnarray}
\label{6a}
\mbox{Re}[\sigma_{\mu\nu}(\omega)] &=& \frac{e^2 \hbar}{2\pi\Omega} 
\sum_{\bf k} \int_{-\infty}^\infty dx \,
\mbox{Tr} \left [ \hat{A}({\bf k};x) \hat{\gamma}_\mu({\bf k},{\bf k}) 
\right. \nonumber \\
&\times& \left. \hat{A}({\bf k};x+\hbar \omega) \hat{\gamma}_\nu({\bf k},
{\bf k}) \right ] \frac{ f(x) - f(x+\hbar\omega)}{\hbar \omega}, \nonumber \\
\end{eqnarray}
where $\hat{A}({\bf k};\omega)$ is the spectral function.  It is a $4\times 4$
matrix whose diagonal elements describe the spectral weight in the
4 superconducting bands.  The electromagnetic vertex functions $\hat{\gamma}$
for $c$-axis transport are,\cite{atkinsonII}
\begin{equation} 
\hat{\gamma} = \left[
	\begin{array}{cccc} v_{z +} & 0 & \alpha T_z & -\beta T_z \\
			    0 & v_{z +} & \beta T_z & \alpha T_z \\
			    \alpha T_z & \beta T_z & v_{z-} & 0 \\
			    -\beta T_z & \alpha T_z & 0 & v_{z-} 
	\end{array}
\right],
\end{equation}
where
\[
\alpha^2 = \frac{1}{2} \left[ 1+\frac{\epsilon_+\epsilon_- + \Delta_{\bf k}^2}
{E_+E_-} \right ],
\]
\[
\beta^2 = \frac{1}{2} \left[ 1-\frac{\epsilon_+\epsilon_- + \Delta_{\bf k}^2}
{E_+E_-} \right ],
\]
are the coherence factors and
\[
T_z = \frac{1}{\hbar}
\frac{\partial t(k_z)}{\partial k_z} \left [ \frac{\xi_1-\xi_2}
{\epsilon_+ - \epsilon_-} \right ]
\]
is the matrix element for the transition.  Furthermore $v_{z\pm}$
are the Fermi velocities
\[
v_{z\pm} = \frac{1}{\hbar}\frac{\partial \epsilon_\pm}{\partial k_z}.
\]

Notice that if the bands $\xi_1$ and $\xi_2$ are nondegenerate, $v_{z\pm}
\propto t_\perp^2$.  In contrast, the off diagonal matrix elements
in $\hat \gamma$ are proportional to $t_\perp$.  Since the diagonal 
matrix elements are for the intraband (or Drude) conductivity, we
can conclude that, for small $t_\perp$, the Drude peak will be small
relative to the non-Drude interband conductivity.  Further, since
$\sigma_c(\omega)$ depends on $\hat \gamma^2$,  the Drude part of
Re $[\sigma_c(\omega)]$ is proportional to $t_\perp^4$, while the
interband part is proportional to $t_\perp^2$.

In the clean limit, the spectral function is $A_{ii}({\bf k};\omega)
= 2\pi\delta(\omega-E_i)$, the intraband contribution to 
$\sigma_{c}(\omega>0)$ vanishes and Eq.~(\ref{6a}) reduces to
\begin{eqnarray}
&&\mbox{Re}\, [\sigma_c (\omega > 0)] = \frac{2\pi e^2 \hbar}{\Omega}
\sum_{\bf k} T_z^2 \nonumber \\
&&\times
\left \{ \alpha^2 \frac{f(E_-) - f(E_+)}{E_+ - E_-} \delta(\hbar \omega
- |E_+ - E_-|) \right . \nonumber \\
&&+
\left . \beta^2 \frac{1-f(E_-) - f(E_+)}{E_+ + E_-} \delta(\hbar \omega
- E_+ - E_-) \right \}.
\label{50}
\end{eqnarray}
This is our basic result.  The first and second terms in the integrand in
(\ref{50}) represent interband transitions of thermally excited quasiparticles
and pair creation of quasiparticles respectively.  We will discuss
these two processes in more detail in the next section.

\section{Results and Discussion}
\label{secIII}

\begin{figure}[t]
\begin{center}
\leavevmode
\epsfxsize \columnwidth
\epsffile{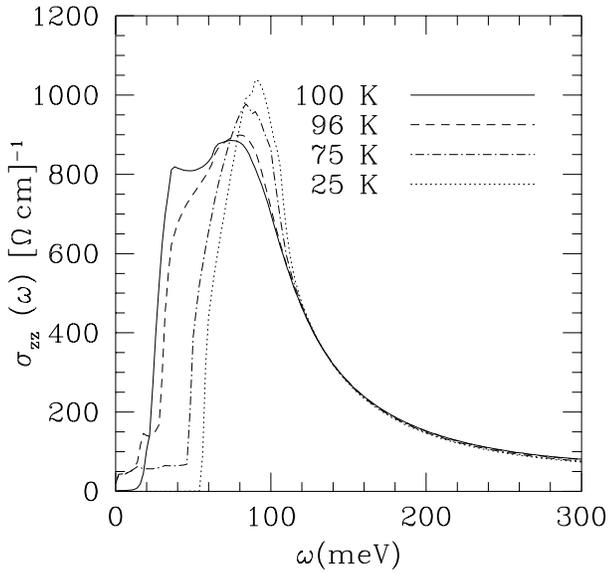}
\caption{The conductivity Re $[\sigma_c(\omega)]$ is shown 
for a $d$-wave gap and for a range of temperatures.
The calculation is in the clean limit, so that the conductivity is
entirely due to interband transitions.  Notice that even though there is
gap in the normal state spectrum (at $T=T_c=100$ K), the superconducting
spectrum is gapless at finite temperature.}
\label{f3}
\end{center}
\end{figure}

\begin{figure}[t]
\begin{center}
\leavevmode
\epsfxsize \columnwidth
\epsffile{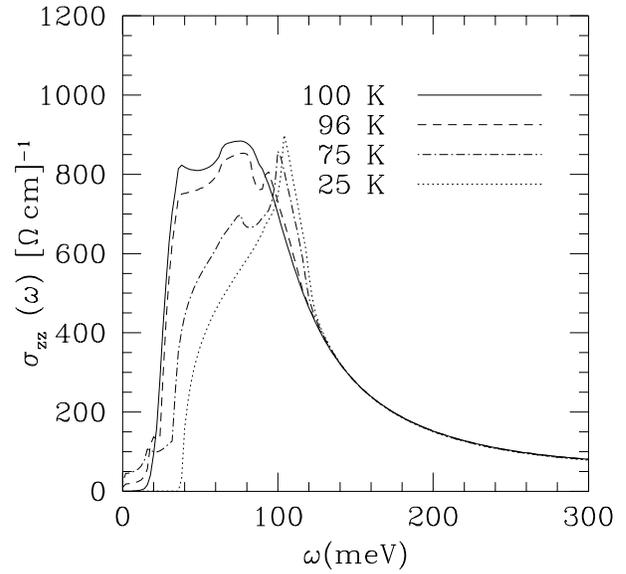}
\caption{The conductivity Re $[\sigma_c(\omega)]$ is shown, as in
Fig.~\protect\ref{f3}, but for an $s$ wave gap.}
\label{f4}
\end{center}
\end{figure}

Equation~(\ref{50}) is integrated numerically and the results are shown in 
Figs.~\ref{f3} and \ref{f4} for $s$ and $d$-wave gaps respectively.
The conductivity is calculated for several temperatures between
$T=0$ K and $T = T_c = 100$ K.  At $T=T_c$ the system is in the normal
state and
\begin{equation}
\mbox{Re }[\sigma_c(\omega)] = \frac{2\pi e^2\hbar}{\Omega} \sum_{\bf k} 
T_z^2 \frac{f(\epsilon_-) - f(\epsilon_+)}{\epsilon_+ - \epsilon_-}
\delta(\hbar \omega - \epsilon_+ + \epsilon_-).
\label{51}
\end{equation}
Equation~(\ref{51}) has the form of a joint density of states since the
integrand is proportional to $\delta(\hbar \omega - \epsilon_+ 
+ \epsilon_-)$.  The thermal factors ensure that the interband transitions
are between filled and empty states.  At low $T$, transitions occur 
between states for which $\epsilon_-({\bf k}) < 0$ and $\epsilon_+({\bf k})
>0$.  

In the high $T_c$ superconductors,
the large frequency range over which $\sigma_c(\omega)$ 
extends is often ascribed
to either a large scattering rate for interplane transitions, or to
the non-Fermi liquid like nature of the ground state.  Here, the frequency
range is of the order of the bandwidth because the energy difference
$\epsilon_+({\bf k}) - \epsilon_-({\bf k})$ 
extends over a wide range of energies.  In Fig.~\ref{f4a}(a) the
energy difference $\epsilon_+ ({\bf k}) - \epsilon_-({\bf k})$ is $\sim 20$ meV
at the point $X$, where the Fermi surfaces are close together, and is
$\sim 375$ meV at the point $Z$.

For our particular choice of parameters, the normal state conductivity
has a finite band gap, which we identify with the 
pseudogap seen in optical conductivity experiments.  The minimum value of  
$\hbar \omega_0 = \epsilon_+({\bf k}) - \epsilon_-({\bf k})$ for which 
$\sigma_c(\omega_0) \neq 0$ is at the point $X$.  Clearly the value of
$\omega_0$ depends on the distance between the two pieces of Fermi
surface.  If the Fermi surfaces cross, then $\omega_0$ vanishes and
arbitrarily low energy excitations are possible.  In this case there
is no pseudogap, and we claim that this describes YBCO in the 
the optimally doped and overdoped cases.

\begin{figure}[t]
\begin{center}
\leavevmode
\epsfxsize \columnwidth
\epsffile{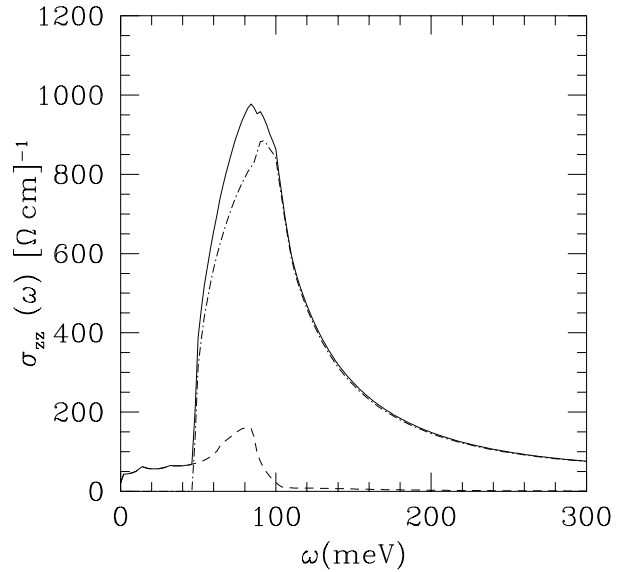}
\caption{The interband conductivity is shown at 75 K (solid curve)
for a $d$-wave gap.  There are two contributions to the conductivity:
a quasiparticle term results from
interband transitions of thermally excited quasiparticles (dashed
curve) and a pair creation term (dot-dashed curve).  The quasiparticle
contribution is gapless and vanishes at $T=0$.  The pair creation
term is gapped.}
\label{f5}
\end{center}
\end{figure}

As the temperature is lowered and the material becomes superconducting,
two types of interband process become possible.  As in the normal state,
quasiparticles which occupy one of the bands may make transitions into
the other band.  This is described by the first term in (\ref{50}) and
the energy required for the transition is $\hbar \omega = 
|E_+({\bf k}) - E_-({\bf k})|$.
Below $T_c$, however, most of the electrons are in
the superconducting condensate, and the largest contribution to the
interband conductivity comes from the creation of quasiparticle pairs,
in which a Cooper pair is broken and one quasiparticle goes into each
of the two bands.  This is described by the second term in (\ref{50}),
and the energy required for the transition is $\hbar \omega = 
E_+({\bf k}) + E_-({\bf k})$.
In Fig.~\ref{f5}, $\sigma_c(\omega)$ is shown as the sum of the
quasiparitcle and pair creation terms.

The two types of interband transition are illustrated in Fig.~\ref{f4a}(b),
and we can make a few comments about Eq.~(\ref{50}) just by inspection
of the figure.  The first is that
the quasiparticle term is not gapped since excitations of arbitrarily
low energy
are available near ${\bf k}$-points where $E_+({\bf k}) = E_-({\bf k})$.  
The probability
for such transitions to occur, however, is strongly 
supressed by the probability that
the lower energy band is initially occupied.  At $T=0$, the thermal
factor 
$
f(E_+) - f(E_-),
$
ensures that the quasiparticle term vanishes.  

A second point we can make is that the pair creation term is gapped
since the energy of pair creation $\hbar \omega = E_+({\bf k})
+E_-({\bf k})$ has, in general, a nonzero minimum value.
The lowest energy pairs which can be created can obviously be 
found by minimizing
$\sqrt{\epsilon_+({\bf k})^2 + \Delta_{\bf k}^2} + 
\sqrt{\epsilon_-({\bf k})^2 + \Delta_{\bf k}^2}$.  For the case where the Fermi
surfaces cross, so that there is a line of ${\bf k}$-values along which
$\epsilon_+({\bf k}) = \epsilon_-({\bf k}) = 0$, the minimum energy is near
$\hbar \omega = 2\Delta_{\bf k}$.  Importantly, this shows that the
pair creation term is gapped for a $d$-wave superconductor
unless the Fermi surfaces happen to cross at $\Delta_{\bf k} = 0$. 

Perhaps the most striking feature of Figs.~\ref{f3} and \ref{f4} is
that the symmetry of the gap does not reveal itself in any obvious
fashion.  It is customary---within a Drude model---to associate a 
gapped frequency dependence with 
an $s$-wave order parameter, and a gapless frequency dependence with
a $d$-wave order parameter.
It is clear from the above discussion, however,
that this cannot be done here.  To summarize our discussion simply,
the interband contribution to the conductivity---which is dominant at
low temperatures---probes the structure of the energy sum
 $E_+ + E_-$ and the energy difference $|E_+ - E_-|$, 
and not of the gap energy $\Delta_{\bf k}$.

This is in contrast, for example, with the single layer model
in which the $c$ axis transport is through diffusive scattering.%
\cite{graf,hirschfeld,rojo}
In this model the conductivity probes the density of states of a 
single layer.\cite{hirschfeld}

\section{Conclusions}
\label{secIV}
For layered superconductors, in which there are more than one type
of layer,  the $c$ axis conductivity is dominated by interband
effects when the interlayer coupling is weak.  In other words, it
is easier for an electron to travel along the $c$ axis by making
a series of interband transitions than by staying within a band.
The intraband (or Drude) and interband conductivities vary as
$t_\perp^4$ and $t_\perp^2$ respectively.

There are two types of interband transition.  The first, interband
transitions of thermally excited quasiparticles, has a gapless spectrum
and probes the joint density of states of the dispersion $E_+({\bf k})
-E_-({\bf k})$.  The second, the excitation of quasiparticle pairs, has
a gapped spectrum and probes the joint density of states of the
dispersion $E_+({\bf k})+E_-({\bf k})$.  
Since neither of these processes probes the
structure of a single band, there is no clear signature of the gap
symmetry in Re $[\sigma_c(\omega)]$.

\section*{Acknowledgements}
Research supported in part by the Canadian Institute for Advanced Research 
(CIAR) and by the Natural Sciences and Engineering Research Council of Canada 
(NSERC).  One of the authors (W.\ A.) is supported in part by the Midwest
Superconductivity Consortium through D.O.E.
grant \# DE-FG-02-90ER45427.

\end{document}